\documentclass[aps,twocolumn,preprintnumbers,longbibliography]{revtex4-1}

\usepackage{graphicx}  % Include figure files
\usepackage{subfigure}
\usepackage{multirow}

\usepackage{fancyhdr}
\usepackage{longtable}
\usepackage{parskip}

\usepackage[T1]{fontenc}
\usepackage{dcolumn}   % Align table columns on decimal point

\usepackage{bm}        % bold math
\usepackage{amsfonts}  % Common math fonts
\usepackage{amsmath}   % Common math functions
\usepackage{amssymb}   % Common math symbols
\usepackage{appendix}

\usepackage[colorlinks=true, filecolor=blue, citecolor=blue,urlcolor=blue]{hyperref}

\usepackage{tikz}      % math illustration lib
\usetikzlibrary{arrows.meta}
\usetikzlibrary{decorations.markings}
\usepackage{xcolor}  %color lib

\newcommand{\eq}{\begin{eqnarray}}
\newcommand{\en}{\end{eqnarray}}

\newcommand{\Evec}{ \vec{E}}

\newcommand{\emt}{{\text{em}}}
\newcommand{\rec}{{\text{rec}}}

\newcommand{\sign}{{\text{sign}}}

\newcommand{\kA}{k_{A'}}
\newcommand{\ka}{k_{a}}

\newcommand{\kapa}{\varkappa_{a}}

\newcommand{\ma}{m_{a}}

\newcommand{\lvec}{\vec{l}}
\newcommand{\xvec}{\vec{x}}

\begin{document}
\preprint{INR-TH-2024-006}
\title{Probing axion-like particles with RF cavities separated by thin barrier}

\author{Dmitry~Salnikov}
\email[\textbf{e-mail}: ]{salnikov.dv16@physics.msu.ru}
%\thanks{corresponding author}
\affiliation{Moscow State University, 119991 Moscow, Russia}
\affiliation{Institute for Nuclear Research, 117312 Moscow, Russia}

\author{D.~V.~Kirpichnikov}
\email[\textbf{e-mail}: ]{dmbrick@gmail.com}
%\thanks{corresponding author}
\affiliation{Institute for Nuclear Research, 117312 Moscow, Russia}

\author{Petr~Satunin}
\email[\textbf{e-mail}: ]{petr.satunin@gmail.com}
\affiliation{Moscow State University, 119991 Moscow, Russia}
\affiliation{Institute for Nuclear Research, 117312 Moscow, Russia}

%\author{Maxim Fitkevich}
%\email[\textbf{e-mail}: ]{fitkevich@phystech.edu}
%\affiliation{Moscow Institute of Physics and Technology,  
% 141701 Dolgoprudny, Russia}
%\affiliation{Institute for Nuclear Research, 117312 Moscow, Russia}

\date{\today}

\begin{abstract}  
We address the Light-Shining-Through-a-thin-Wall (LSthinW) laboratory setup to estimate the sensitivity of axion-like particle (ALP) detection using two radio-frequency (RF) cavities immersed in a static magnetic field. We analytically evaluate the asymptotic sensitivity in the off-shell regime for the lowest electromagnetic pump modes. We show that a sufficiently thin wall separating can lead to the improved  sensitivity of the pure laboratory probes of ALP in the mass range $10^{-6}~{\rm eV} \lesssim m_a \lesssim 10^{-4}~{\rm eV}$.
\end{abstract}

\maketitle 

\section{Introduction. } 

Feebly-interacting particles (FIPs) \cite{Antel:2023hkf} are hypothetical light particles which extremely 
weak interact with the Standard Model (SM) fields and may consist of the dark matter component of the Universe 
or can be a mediator between Dark Matter and  SM sector. The well motivated  FIPs are 
photophilic: pseudoscalar axion-like-particle (ALP) \cite{Svrcek:2006yi, Arvanitaki:2009fg,Berezhiani:1989fp,Sakharov:1994id} and vector 
dark photon (DP) \cite{ Okun:1982xi, Holdom:1985ag}. Both of them can be tested with their interaction with 
Standard Model photons.

A typical strategy for probing FIPs (ALPs and DPs) implies both their production and detection in a laboratory, and
usually called Light-Shining-through-Wall (LSW) experiments 
\cite{Sikivie:1983ip,Anselm:1985obz,VanBibber:1987rq, Hoogeveen:1992nq,Salnikov:2020urr,Gao:2020anb}. 
The  LSW setups consist of 
two cavities separated by a non-transparent 
wall for ordinary photons.  The FIPs are produced in the first cavity by a cavity electromagnetic mode (in case of DPs) or by an interaction of two electromagnetic field components (in case of ALPs). 

Generated FIPs can pass through the wall and convert back to photons in the 
detection cavity.  In the present paper, we  
focus on the pure laboratory LSW setup for probing ALPs that consists of two radio frequency (RF) cavities 
separated by a thin barrier. We also show that regarding expected reach can be comparable with the  the typical LSW laser-based experimental limits~\cite{Ehret:2010mh, OSQAR:2015qdv}.  Typically, LSW RF cavity experiments, such as 
CROWS~\cite{Betz:2013dza}, DarkSRF~\cite{Romanenko:2023irv}, imply that the FIPs emitter and receiver cavities are  divided by a 
distance that is comparable to the typical size of the detector. This setup is mostly sensitive to FIPs 
mass less than the emitter mode frequency since the heavier FIPs are virtual and their density 
exponentially decrease with the distance between cavities.

Recently the authors of \cite{Berlin:2023mti} 
proposed an idea that a sufficiently thin wall between cavities significantly improves the DP sensitivity of 
in the region of the sufficiently large DP mass, such that it can be considered as off-shell state (see also \cite{Salnikov:2024ovk}). %(see also \cite{Salnikov:2024ovk} for another  cavity configuration). 
In this paper, we address the thin wall idea for probing ALPs in LSW experiments.
In particular, we argue that employing
the sufficiently low driven emitter frequency $\omega\lesssim 1~\,\mbox{GHz}$ and thin barrier between cavities $d \ll 1/\omega$,
one can achieve a large density  of the ALP field. As a result this leads to the improved  sensitivity in 
the off-shell ALP mas region $m_a \gtrsim \omega$, if one compares it with the LSW purely laboratory 
  CROWS~\cite{Betz:2013dza}  experiment.
  We also show that the expected reach of suggested  LSthinW setup is complementary to the laser-based laboratory 
  limits of   ALPS-I~\cite{Ehret:2010mh} and OSQAR~\cite{OSQAR:2015qdv} experimental facilities. 

  This paper is organized as follows. In Sec.~\ref{SectLagr} we discuss the ALP electrodynamics. In Sec.~\ref{SectSetup} we derive main formulas in order to estimate the 
  sensitivity of the LSthinW setup to probe ALPs. In Sec.~\ref{SectResults} we discuss the numerical results for the expected sensitivity. In  Appendices \ref{AppSectQuantum} and \ref{AppInt} we discuss some useful formulas. 

\section{The ALP electrodynamics \label{SectLagr}}

The Lagrangian describing coupling of  ALPs and Standard model photons reads
\begin{equation}\label{Lagrangian}
\mathcal{L} \supset-\frac{1}{4} F_{\mu\nu}F^{\mu\nu}+\frac12\,\partial_\mu a\, \partial^\mu a - \frac12\,m_a^2 a^2 -\frac{1}{4}\,g_{a\gamma\gamma}\,a\,F_{\mu\nu}\tilde{F}^{\mu\nu}\;,
\end{equation}
where $F_{\mu\nu}$ is the electromagnetic tensor and 
$\tilde{F}^{\mu\nu}=\frac{1}{2}\epsilon^{\mu\nu\alpha\beta}F_{\alpha\beta}$ is its dual, $a$ denotes the 
ALP field,  and $g_{a\gamma\gamma}$ is dimensional ALP-photon coupling. 

%A typical strategy for probing ALPs implies both their production and detection in a laboratory, and usually called Light-Shining-through-Wall (LSW) experiments \cite{Sikivie:1983ip,Anselm:1985obz,VanBibber:1987rq, Hoogeveen:1992nq}. The LSW setups consist of two cavities separated by a non-transparent wall for ordinary photons.  The ALPs are produced in the first cavity by interaction of electromagnetic field components. Generated ALPs can pass through the wall and convert back to photons in the detection cavity. Recently, several proposals with LSW radio cavities  appeared in the literature  radio frequency (RF)   cavities~\cite{Gao:2020anb,Salnikov:2020urr, Bogorad:2019pbu}. In this paper we compare  different LSW cavity setups including modification of the CROWS \cite{Hoogeveen:1992nq,Betz:2013dza}. 

%\section{Axion electrodynamics.}
The Lagrangian (\ref{Lagrangian}) implies that the equations 
for the ALP field and SM photon in vacuum read respectively,
\begin{align}
(\partial_\mu \partial^\mu +m_a^2)\,a=g_{a\gamma\gamma}(\vec{E} \cdot \vec{B})\;,
\label{Klein_Gordon_eqn2} \\
    (\vec{\nabla}\cdot \vec{E}) =  \rho_a\;, \quad\quad 
    [\vec{\nabla}\times \vec{B}] = \dot{\vec{E}} + \vec{j}_a\;, \label{Maxwell1}
\end{align}
where the density of charge $\rho_a$ and current $\vec{j}_a$, induced by the ALP, %these quantities 
read respectively as follows
\begin{equation}
\label{GeneralRhoaAndJa1}
    \rho_a = -g_{a\gamma\gamma} (\vec{\nabla}a\cdot \vec{B})\;, \quad\vec{j}_a = g_{a\gamma\gamma}( [\vec{\nabla}a\times \vec{E}] + \dot{a}\vec{B} )\;.
\end{equation}
We note that Eq.~(\ref{Klein_Gordon_eqn2}) describes
the ALP production by the typical
combination of electromagnetic fields $\propto (\vec{E} \cdot \vec{B}) $ in its 
right-hand side. Contrary, both Eqs.~(\ref{Maxwell1}) and  (\ref{GeneralRhoaAndJa1}) can
be associated with the ALP conversion in the strong magnetic field that results in the 
generation of a resonantly enhanced signal EM mode in the detector.  

\begin{figure*}[tbh!]\centering
\includegraphics[width=0.37\linewidth]{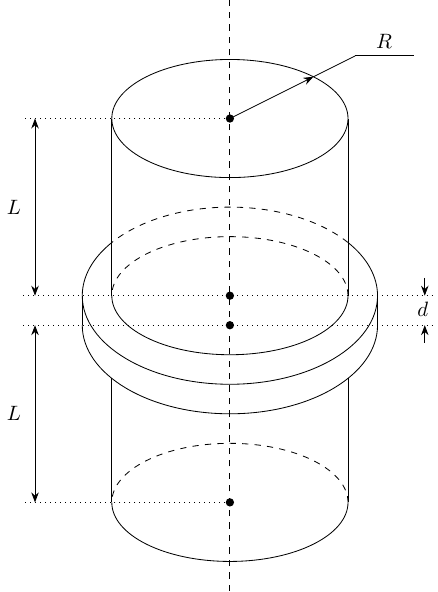}
\includegraphics[width=0.62\linewidth]{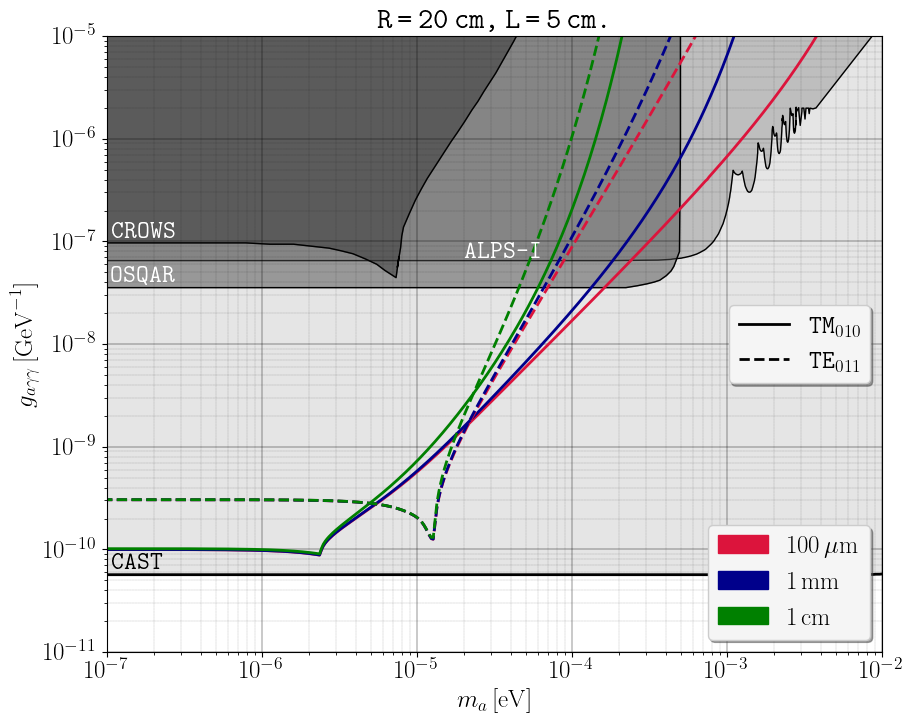}
\caption{{\it Left panel}: The typical scheme of LSthinW setup. {\it Right panel}: The setup sensitivity to ALP parameter space for TM$_{010}$ (solid lines) and TE$_{011}$ (dashed lines) modes. The green, blue, and red lines correspond to the barrier thickness of
$d= 1~\mbox{cm}$, $d= 1~\mbox{mm}$, and $d= 100~\mu\mbox{m}$ respectively, the radius and length of the cavities are chosen to be $R=20~\mbox{cm}$ and $L=5~\mbox{cm}$, respectively. Emitter power is taken as $10$ kW.
The pure laboratory bounds of CROWS~\cite{Betz:2013dza},  OSQAR~\cite{OSQAR:2015qdv}, and ALPS-I~\cite{Ehret:2010mh} are shown by gray regions.  The extraterrestrial helioscope limit of CAST \cite{CAST:2017uph} 
is shown by a solid black line. All limitations correspond to 95\% CL.
}
\label{fig:1}
\end{figure*}

\section{Light-Shining-through thin Wall setup \label{SectSetup}} 
In this section, we discuss  the following benchmark experimental setup consisting of two RF cavities. 
We consider two equal cylindrical cavities of a radius $R$ and length $L$, located coaxial end-by-end and separated 
%In particular, we consider the cylindrical receiver that is coaxial and separated from the emitter's endcap
by a thin barrier of a length $d \ll L$, see Fig.~\ref{fig:1}, left panel. %, where $R$ is a typical radius of, moreover
We compare two cases of the lowest pump modes, namely ${\rm TM}_{010}$ and ${\rm TE}_{011}$. 
Both %emitter and receiver 
cavities  are immersed in a static 
magnetic field $B_{\rm ext} =10~\mbox{T}$. The field is aligned along the cavities' axis in the case of the ${\rm TM}_{010}$ pump mode, and perpendicular to the axis in the case of the ${\rm TE}_{011}$ pump mode. %$z$-axis

\paragraph{The ALP emission:} the emitter cavity is pumped by an electromagnetic  mode with electric field, given by $\Evec_{\emt} (\xvec) = E^\emt_0 
\vec{\mathcal{E}}^\emt(\xvec) e^{-i \omega t}$, where $\omega$ is a driven frequency of the pump mode, $E_0^\emt =3~\mbox{MV}/\mbox{m} =0.01~\mbox{T}$ is a magnitude of the driven emitter electric field,
$\vec{\mathcal{E}}^\emt(\xvec)$ is a dimensionless electric field that is normalized as
$$ 
\int\limits_{V_\emt} d^3  \xvec \, |\vec{\mathcal{E}}^\emt(\xvec)|^2 = V_\emt. 
$$

The emitter power can be expressed as
\begin{equation}\label{emmiter}
   P_{\rm em} = \dfrac{\omega}{Q_{\rm em}} \dfrac{1}{2} |E^{\rm em}_0|^2 V_{\rm em},
\end{equation}
and for the considered values of the pump mode amplitude and the characteristic dimensions of the resonator, $R = 20\ {\rm cm},\ L = 5\ {\rm cm}$, it is approximately $P_{\rm em} \simeq 10\ {\rm kW}$.

In the cases of TM$_{010}$ and TE$_{011}$ modes, corresponding dimensionless electric field read
\begin{align}
     {\rm TM_{010}} & : \vec{\cal E}^{\rm em}(\vec x) = {\vec e}_z   J_0(x_{01}\rho/R) / |J'_0(x_{01})|, \notag \\
 {\rm TE_{011}} &: \vec{\cal E}^{\rm em}(\vec x) = {\vec e}_\varphi  \sqrt{2} J_1(x_{11}\rho/R) \sin(\pi z/L)/|J_1'(x_{11})|, \notag
\end{align}
where $x_{01}$ and $x_{11}$ are the first zeros of Bessel functions $J_0(x)$ and $J_1(x)$, respectively.

%(see Fig.~\ref{fig:DistDesign});
%As a result, the right hand side of the Eq.~(\ref{Klein_Gordon_eqn2}) yields the typical source combination of the field  in the following form
%\begin{equation}\label{Source}
%     g_{a\gamma\gamma}E^{\rm em}_0 B_{\rm ext} \mathcal{E}^\emt_z (\xvec)\cdot e^{-i\omega t} \;,
%\end{equation}
%therefore

The Eq.~\eqref{Klein_Gordon_eqn2} for produced ALP field implies the following solution,
\begin{align}
\label{Produced_ALPs}
   a(t,\vec{x}) = g_{a\gamma\gamma}\, E^{\rm em}_0 B_{\rm ext} \!\!\! \int\limits_{V_\emt}\!\! d^3 \vec{x}' \mathcal{E}^\emt_B(\vec{x}')  
   \dfrac{e^{ik_a|\vec{x} - \vec{x}'| - i\omega t}}{4\pi|\vec{x} - \vec{x}'|}, 
\end{align}
where ${\cal E}^{\rm em}_B$ is a projection of pump mode electric field on the direction of external constant magnetic field ($z$-axis in the case of TM$_{010}$ mode and $x$-axis in the case of TE$_{011}$ mode), $B_{\rm ext}$ is a characteristic magnetic field which is chosen to be $B_{\rm ext} = 10 \, {\rm T}$, $k_a=\sqrt{\omega^2 - m_a^2}$ are typical momenta of the produced ALPs; the integration is performed over the emitter volume $V_{\rm em}$. 
One replaces  $i k_a$ with $-\kapa=-\sqrt{m_a^2 - \omega^2}$ in Eq.~(\ref{Produced_ALPs}) for 
$m_a \gtrsim \omega$.  

\paragraph{The ALP detection:} Now we  discuss  the signal induced by the ALP field in the resonant {\it receiver} 
cavity which is a copy of the emitter cavity, so we would search for exactly the same mode 
$\vec{\mathcal{E}}^\emt(\xvec) = \vec{\mathcal{E}}^\rec(\xvec)$ which is expected to be grown  resonantly. %Mathematically, one  integrates the second equation of  (\ref{Maxwell1}) over the receiver cavity volume with the normalized receiver cavity mode,
%and introducing dissipation decrement $\frac{Q_\rec}{\omega}$ by hands (see ~ for details),
 In particular, the resonant amplitude of the receiver is characterized by the 
 term~\cite{Gao:2020anb, Bogorad:2019pbu}   
%A resonant generation of electromagnetic modes in the detecting cavity caused by the axion-induced current (see Eqs.~(\ref{GeneralRhoaAndJa1}) for detail). 
%In particular, the typical magnitude of the resonantly induced signal in the receiver is characterized by the  overlapping factor~\cite{Gao:2020anb, Bogorad:2019pbu}
\begin{equation}\label{Amplitude}
    G = -\dfrac{Q_{\rm rec}}{\omega_s}\cdot \dfrac{1}{V_{\rec}}\int\limits_{V_{\rec}} d^3 x\, 
    \vec{\mathcal{E}}^*_{\rec} (\xvec) 
    \vec{j}_a (\xvec)\;,
\end{equation}
where $Q_{\rm rec}$ is a quality factor for the receiver eigenmode and $V_{\rm rec}$ is the volume of the receiver 
cavity,  $\omega_s \simeq \omega$ is a frequency of the receiver signal eigenmode, and $\vec{\mathcal{E}}_\rec (\xvec) $ is 
a dimensionless signal eigenmode. The   expression  Eq.~(\ref{Amplitude})  reads
\begin{equation}\label{Detecting_Integral}
   G =  -i g_{a\gamma\gamma} \frac{Q_\rec B_{\rm ext}}{V_{\rm rec}}\! \!\! \int\limits_{V_{\rm rec}}\! \! \!d^3 x\,   \mathcal{E}^{\rec *}_B (\xvec)  a(\vec{x})\;,
\end{equation}
where $a(\vec{x})$
is a stationary part of Eq.~(\ref{Produced_ALPs}). 
The typical signal power reads
\begin{equation}\label{Signal_Power}
    P_{\rm signal} \!  = \!\dfrac{\omega}{Q_{\rm rec}} \dfrac{1}{2} |G|^2 V_{\rm rec}.
\end{equation}

Now let us estimate the sensitivity numerically as maximum output in the receiver cavity that is  given by the Dicke radiometer equation,~\cite{Dicke:1946glx},
\begin{equation}
 \mathrm{SNR} =\frac{P_{\mathrm{signal}}}{P_\mathrm{noise}}\cdot  \sqrt{t\Delta \nu}\;,
 \label{SNR1}
\end{equation}
where $t$ is an integration time for a signal, $\Delta \nu$ is its bandwidth and $P_{\rm noise}$ is a power of thermal 
noise which can be estimated as $P_{\rm noise}\simeq T\Delta \nu$ in the limit $\omega \ll T$, where $T\simeq 1.5\,\mbox{K}$ is the typical 
temperature of the receiver. We consider the narrowest possible bandwidth of a pump generator, which can be as small as 
$\Delta \nu  \simeq 1/t$ (see e.~g.~Refs.~\cite{Bogorad:2019pbu, Gao:2020anb} and references therein).  The quality factor is chosen to be 
$Q_{\rec}\simeq 10^{5}$ throughout the paper. Moreover, in the numerical estimate 
we conservatively set  integration time to be 
$t\simeq 8.6\cdot10^{4}~\mbox{s} \simeq 1~\mbox{day}$.  
We address the quantum description of the signal in Appendix~\ref{AppSectQuantum}.

Let  $\xvec'$ and $\xvec$ be the  coordinates associated with  the emitter and receiver frame respectively. 
In addition, let $\lvec$ be the vector linking the origins of the  frames of both 
emitter and receiver for the coaxial design of the experimental setup. In this case  one has a simple link 
between them~$\xvec' = \lvec + \xvec$.  The vector linking the cavity frames would be  then $\lvec = (0,0,l_z)$. 
For that notation the distance between closest endcaps of the cavities is
$d = l_z -L$. 

In the following we rewrite $G$ via  
%Let us introduce 
the dimensionless form-factor ${\cal G}$
$$
G = {\cal G} \, {g_{a\gamma\gamma}^2E_0^{\emt}B_{\rm ext}^2Q_\rec V_{\emt}\omega},
$$
where the expression for ${\cal G}$ reads
\begin{equation}
    {\cal G} \! = \! \dfrac{1}{\omega} \! \! \int\limits_{V_{\rm rec}}\! \!\! \dfrac{d^3 x}{V_{\rm rec}} \!\!\int\limits_{V_{\rm em}} \!\! \dfrac{d^3x'}{V_{\rm em}} \, \mathcal{E}^{\rec *}_B (\xvec)   \mathcal{E}_B^\emt (\xvec')    \dfrac{e^{ik_a|\Vec{x} - \vec{x}'-\lvec|}}{4\pi|\vec{x} - \vec{x}' - \lvec|}.
    \label{DmnslsFormFactorDef1}
\end{equation}

The corresponding integral can be evaluated analytically, as demonstrated in \cite{Salnikov:2024ovk}. The details are provided in Appendix \ref{AppInt}.

Finally, by using Eqs.~(\ref{Signal_Power}) and (\ref{SNR1}) we obtain the expected sensitivity,
\begin{equation}
\label{master}
    g_{a\gamma\gamma} = \left[ \dfrac{ T \, {\rm SNR}  }{B_{\rm ext}^4 P_{\rm em} Q_{\rm em}Q_{\rm rec}V_{\rm em}\!V_{\rm rec}\omega^2 |{\cal G}|^2 t} \right]^{1/4},
\end{equation}
where the signal-to-noise ratio is chosen to be ${\rm SNR} \simeq 1.65$ (one-side 95\% CL) for the sensitivity estimate.

\section{Results and Discussion \label{SectResults}}
We have analytically calculated the asymptotic behavior of the geometric form factor $G$ and the expected sensitivity for large masses ($m_a d \gg 1$) in a radio-frequency cavity Light-Shining-through-a-thin-Wall setup for axion-like particles (ALPs), considering the two lowest electromagnetic modes, TM$_{010}$ and TE$_{011}$, in the presence of an external constant magnetic field.

The asymptotics read:
\begin{enumerate}
    \item TM$_{010}$-mode:
\begin{align}
    {\cal G} = \ & (2\pi)^{-1} \omega^{-1} R^{-2}L^{-2} m_a^{-3} \exp(m_a d), \\
    %= \ & (2\pi x_{01})^{-1} R^{-1} L^{-2} m_a^{-3} \exp(m_a d), \\
    g_{a\gamma\gamma} 
    = \ & \left[ \dfrac{T \,{\rm SNR}}{B^4_{\rm ext} P_{\rm em} Q_{\rm rec} Q_{\rm em}  t}  \right]^{1/4} \times (2L)^{1/2} \times \notag \\
    & \times m_a^{3/2} \times \exp\left( \dfrac{m_ad}{2} \right).
\end{align}
\item  TE$_{011}$-mode:
\begin{align}
    {\cal G} = \ &  \pi \cdot (2\omega)^{-1} R^{-2} L^{-4} m_a^{-5} \exp(m_a d), \notag \\
    %= \ & 2^{-1} R^{-2} L^{-3} m_{a}^{-5}  \exp(m_a d), \quad ({\rm for} \ R \gg L), \\
    g_{a\gamma\gamma}  = \ &  \left[ \dfrac{T \,{\rm SNR}}{B^4_{\rm ext} P_{\rm em} Q_{\rm rec} Q_{\rm em}  t}  \right]^{1/4} \times \left( \dfrac{2}{\pi^2} \cdot L^3 \right)^{1/2} \times \notag \\
    & \times m_a^{5/2} \times \exp\left( \dfrac{m_ad}{2} \right).
\end{align}
\end{enumerate}

{\setlength{\tabcolsep}{0.5em}
\renewcommand{\arraystretch}{2}% for the vertical padding
\begin{table}[!ht]
 %   \color{blue}
    \centering
    \begin{tabular}{ c c c }
    \hline
    \hline
    Mode  & Geometry prefactor &  Mass dependency  \\
    \hline
    TM$_{010}$ &  $L^{1/2}$ & $m_a^{3/2} \times \exp\left(m_a d/2\right)$ \\
    TE$_{011}$ &  $L^{3/2}$ & $m_a^{5/2} \times \exp\left(m_a d/2\right)$ \\
    \hline
    \end{tabular}
    \caption{Asymptotic behaviour of $g_{a \gamma \gamma}$ at large masses for both TM$_{010}$ and TE$_{011}$ cavity pump modes.}
    \label{tab:results}
\end{table}
}

In Fig.~\ref{fig:1} (right panel) 
we show the setup sensitivity to ALP parameters $g_{a\gamma\gamma}$ 
and $m_a$. The expected reaches are shown for the barrier thickness in the range
$100~\mu\mbox{m} \lesssim d \lesssim 1~\mbox{cm}$.
For given benchmark design of the LSthinW setup ($R \simeq  20~\mbox{cm}$, 
$L \simeq 5~\mbox{cm}$) 
the resonant  enhancement of the ALP sensitivity $g_{a\gamma \gamma} \sim  10^{-10}~\mbox{GeV}^{-1}$ can be achieved for the 
typical  masses   $  \ma \simeq \omega \simeq  10^{-6}~\mbox{eV} \div 10^{-5} \, {\rm eV}$. 
The expected reach for the sufficiently small  ALP masses, $\ma \ll  \omega$, 
 can be at the level of     $g_{a\gamma \gamma} \lesssim   10^{-10}~\mbox{GeV}^{-1}$.

 For the off-shell region of the ALP masses, $m_a \gtrsim \omega$ one can achieve the 
 enhancement of the sensitivity due to the reducing the cavity barrier thickness.  
In particular, Fig.~\ref{fig:1} shows that the smaller the distance between  the cavities, 
$d$  the smaller the slope of the expected reach curve. The exact dependency on mass and geometry prefactor for two types of modes is presented in Table \ref{tab:results}. It is worth noting that the dependence for the TM$_{010}$ mode exhibits a smaller slope compared to the TE$_{011}$ mode and is more preferable for the off-shell regime. The setup with the benchmark 
thickness of the wall at the level of $d\simeq 100~\mu\mbox{m}$ can rule out the typical masses 
of ALP below $m_a \lesssim 10^{-4}~\mbox{eV}$ (in the case TM$_{010}$ pump mode).  Moreover, the regarding region is 
complementary  to the bounds of purely laboratory LWS setup, which are  based on lasers. 

In order to conclude this section, we emphasize that one can employ the copper based cavities 
with typical skin depth of the order of $\delta\simeq~2\mu\mbox{m}$ for 
$\omega \simeq 1~\mbox{GHz}$. This conservatively  implies that  the typical barrier thickness can be as large as 
$d \gtrsim 30 \delta \simeq 60~\mu\mbox{m}$, so that our benchmark value $d \simeq 100~\mu\mbox{m}$ is a fairly 
reasonable wall thickness.   

The possible extension of the present work can be addressed to a probing of millicharged and 
axion-like  particles in LSthinW setup by varying  the cavity design, 
magnetic field orientation and the specific choice of the pump mode of the emitter.  
We leave these tasks for future study. 

The code for numerical calculations and drawing figures can be found here \footnote{\href{https://github.com/salnikov-dmitry/lsw}{https://github.com/salnikov-dmitry/lsw}}.

\section{Acknowledgements}
We would like to thank Maxim Fitkevich, Dima Levkov, Aleksandr Panin,  Leysan Valeeva and Alexey Zhevlakov for 
fruitful discussions.   This work  is supported by  RSF  grant no. 21-72-10151.

\appendix
\section{Quantum description of the signal \label{AppSectQuantum}}
Remarkably, that by employing Schwinger-Keldysh formalism~\cite{Trunin:2022dvg,Kopchinskii:2023whq} one can link  
the coherent transition   probability $P^{\rm coh}_{\gamma \to a \to \gamma}$ with the averaged typical number of photon quanta 
in  the receiver  $\langle  N_{{\rm TM}_{010}} \rangle =\langle  i| \hat{N}_{{\rm TM}_{010}} |i \rangle = \langle i| \hat{a}^\dagger_{{\rm TM}_{010}}  
\hat{a}_{{\rm TM}_{010}} |i \rangle$, where the initial state $|i \rangle$ implies the time evolution from $t_i$ to $t_f$ and back to $t_i$
$$
\langle  N_{{\rm TM}_{010}} \rangle  = \int df' \int df'' \langle i |\hat{S}^\dagger | f' \rangle \langle f' | \hat{N}_{{\rm TM}_{010}} | f'' \rangle  \langle f'' |\hat{S} | i \rangle,
$$ 
where $\hat{S}$ is an S-matrix operator~\cite{Kopchinskii:2023whq} for ALP-photon interaction, 
$|f'\rangle$ and $ |f''\rangle$  represent the full set of states in theory at the moment $t_f$.
Only $ |f'\rangle = |f'\rangle$ states yield nonzero outcome, such that 
$\langle  N_{{\rm TM}_{010}} \rangle =  P^{\rm coh}_{\gamma \to a \to \gamma}$,  implying a few mode 
quanta in a steady regime $ \langle N_{{\rm TM}_{010}} \rangle \lesssim 1$. In addition,  we note that 
the straightforward analytical calculation reveals a well known result~\cite{Bogorad:2019pbu} for which the  
classical number of signal photons in the emitter 
$N_{\rm TM_{010}} = Q_{\rec} P_{\sign }/\omega^2$ is 
$N_{\rm TM_{010}} = \langle  N_{{\rm TM}_{010}} \rangle$. 

\section{Analytical evaluation of the geometric form-factor \label{AppInt}}

In this section, we present the evaluation of the integrals used in the calculation of the form factor~(\ref{DmnslsFormFactorDef1}) asymptotics. We use the Fourier representation of the spherical wave,
\begin{equation}
    \dfrac{e^{ik_a|\vec x-\vec x' - \vec l|}}{4\pi|\vec x - \vec x' - \vec l|} = \int\dfrac{d^3k}{(2\pi)^3} \dfrac{e^{-i\vec k(\vec x- \vec x ' - \vec l)}}{k^2 - (k_a + i\varepsilon)^2}.
\end{equation}

The momentum space is considered in the cylindrical coordinate system, $\vec{k} \to (k_\rho, \varphi_k, k_z)$, $d^3k = k_\rho dk_\rho d\varphi_k dk_z$.
Performing a change in the order of integration, we analytically calculated integrals over $x$, $x'$ and an angle in momentum space $\varphi_k$ in  all geometrical and pump modes cases. 

Integrating over spaces and momentum angles $\varphi, \varphi', \varphi_k$:
\begin{align}
    &\int_0^{2\pi}d\varphi \int_0^{2\pi}d\varphi' \int_0^{2\pi}d\varphi_k \, f_0(\vec{x};\vec{x}';\vec{k})   \\
    &=(2\pi)^3 \times J_n(k_\rho\rho)J_n(k_\rho\rho') \times f_1(\rho,z;\rho',z;k_\rho,k_z),
    \nonumber
\end{align}
where $n = 0$ in the case of TM$_{010}$ mode and $n=1$ in the case of TE$_{011}$ mode.

Integrating over $z$ and $z'$ for the TM$_{010}$ mode reads:
\begin{align}
    &\int_0^L dz \int_0^L dz' \, f_1(\rho,z;\rho',z;k_\rho,k_z)  \\
    &= L^2\times \left[ \dfrac{\sin(\frac{k_zL}{2})}{\frac{k_zL}{2}}\right]^2 \times f_2(\rho; \rho'; k_\rho, k_z), \nonumber
\end{align}
and 
\begin{align}
    &\int_0^L dz \int_0^L dz' \, f_1(\rho,z;\rho',z;k_\rho,k_z) \\
    &= \left(\dfrac{L}{2}\right)^2\times \left[ \dfrac{ \pi \cos(\frac{k_zL}{2})}{\left(\frac{k_zL}{2}\right)^2 - \left(\frac{p_zL}{2}\right)^2} \right]^2 \times f_2(\rho; \rho'; k_\rho, k_z),
     \nonumber
\end{align}
for the TE$_{011}$ mode, where $p_z = \pi/L$ is axial eigenmomentum.

Integrating over $\rho$ and $\rho'$:
\begin{align}
    & \int_{0}^{R} \rho d\rho  \, f_2(\rho;\rho';k_\rho,k_z)  \\
    &=R^2 \times \dfrac{p_\rho R J'_n(p_\rho R) J_n(k_\rho R)}{(k_\rho R)^2 - (p_\rho R)^2} \times f_3(\rho'; k_\rho, k_z), \nonumber
\end{align}
where $p_\rho = x_{n1}/R$ is radial eigenmomentum.

In the case of adjacent geometry, we performed integration for the cylinder twice (over both $\rho$ and $\rho'$).

The remaining two-dimensional integral in momentum space can be reduced by contour integration over $k_\rho$ to a one-dimensional integral according to Jordan's lemma:
\begin{align}\label{k_rho_integration}
&\int^{+\infty}_{0} k_\rho dk_\rho \, \dfrac{J^2_n(k_\rho R)}{[(k_\rho R)^2 - (p_\rho R)^2]^2} \cdot \dfrac{1}{k_\rho^2 - (q + i\varepsilon)^2} \nonumber  \\
&= \dfrac{\pi i}{2} \left[\dfrac{H^{(1)}_n(qR)J_n(qR)}{[(qR)^2 - (p_\rho R)^2]^2} - \dfrac{iY_n(p_\rho R)J'_n(p_\rho R)}{2(p_\rho R)[(qR)^2 - (p_\rho R)^2]} \right]  \nonumber \\
& =\dfrac{\pi i}{2} \left[ F(q) - \tilde{F}(q) \right],
\end{align}
where $q^2 = \omega^2 - k_z^2$ and
\begin{align}\label{k_rho_integration}
F(q) &= \dfrac{H^{(1)}_n(qR')J_n(qR)}{[(qR)^2 - (p_\rho R)^2]^2},\\
\tilde{F}(q) &= \dfrac{\lim\limits_{q \to p_\rho}[\{(qR)^2 - (p_\rho R)^2\}\times F(q)]}{(qR)^2 - (p_\rho R)^2}.
\end{align}

The exact analytical asymptotic can be obtained by the approximations under conditions $\varkappa_{A'}L \gg 1$:
\begin{equation}
    \left[ \dfrac{\sin\left(\dfrac{\varkappa_{A'}Lx}{2}\right)}{\dfrac{\varkappa_{A'}Lx}{2}}\right]^2 \simeq \dfrac{2\pi}{\varkappa_{A'} L} \delta (x),
\end{equation}
\begin{equation}
    \left[ \dfrac{ \pi  \cos\left(\dfrac{\varkappa L x}{2}\right)}{\left(\dfrac{\varkappa L x}{2}\right)^2 - \left(\dfrac{\pi}{2}\right)^2} \right]^2 \simeq \dfrac{4\pi}{\varkappa_{A'} L} \delta (x),
\end{equation}
which reduces the remaining one-dimensional integral over the $k_z$.

\bibliography{bibl}

\end{document}